\documentclass[preprint,12pt]{elsarticle}

\usepackage{amssymb}
\usepackage{amsmath}
\usepackage[T1]{fontenc}
\usepackage{graphicx}
\usepackage{algorithm}
\usepackage{algorithmic}
\usepackage{booktabs}
\usepackage{subcaption}
\usepackage{threeparttable}
\usepackage[hidelinks]{hyperref}
\usepackage{color}
\usepackage{url}
\usepackage{mwe}
\usepackage{multirow}
\usepackage{array}
\usepackage{caption}
\usepackage[misc]{ifsym}
\usepackage{float}
\usepackage{booktabs} 
\usepackage{makecell}

\usepackage{pifont}
\newcommand{\cmark}{\ding{51}}
\newcommand{\xmark}{\ding{55}}

\journal{Knowledge-Based Systems}

\begin{document}

\begin{frontmatter}



\title{WOCD: A Semi-Supervised Method for Overlapping Community Detection Using Weak Cliques}


\author[label1]{Shaozhen Ma} 
\author[label2]{Hanchen Wang}
\author[label1]{Dong Wen}
\author[label1]{Wenjie Zhang}
\author[label1]{Wei Huang\corref{cor1}}
\ead{w.c.huang@unsw.edu.au}
\author[label3]{Ying Zhang}
\cortext[cor1]{Corresponding author.}

\affiliation[label1]{organization={University of New South Wales},
            addressline={Kensington}, 
            city={Sydney},
            postcode={2052}, 
            state={NSW},
            country={Australia}}
\affiliation[label2]{organization={University of Technology Sydney},
            addressline={Ultimo}, 
            city={Sydney},
            postcode={2007}, 
            state={NSW},
            country={Australia}}
\affiliation[label3]{organization={Zhejiang Gongshang University},
            addressline={Xiasha}, 
            city={Hangzhou},
            postcode={310018}, 
            state={Zhejiang},
            country={China}}

\begin{abstract}
Overlapping community detection (OCD) is a fundamental graph data analysis task for extracting graph patterns.
Traditional OCD methods can be broadly divided into node clustering and link clustering approaches, both of which rely solely on link information to identify overlapping communities.
In recent years, deep learning-based methods have made significant advancements for this task. 
However, existing GNN-based approaches often face difficulties in effectively integrating link, attribute, and prior information, along with challenges like limited receptive fields and over-smoothing, which hinder their performance on complex overlapping community detection. 
In this paper, we propose a Weak-clique based Overlapping Community Detection method, namely WOCD, which incorporates prior information and optimizes the use of link information to improve detection accuracy. 
Specifically, we introduce pseudo-labels within a semi-supervised framework to strengthen the generalization ability, making WOCD more versatile.
Furthermore, we initialize pseudo-labels using weak cliques to fully leverage link and prior information, leading to better detection accuracy.
Additionally, we employ a single-layer Graph Transformer combined with GNN, which achieves significant performance improvements while maintaining efficiency. 
We evaluate WOCD on eight real-world attributed datasets, and the results demonstrate that it outperforms the state-of-the-art semi-supervised OCD method by a significant margin in terms of accuracy. 
\end{abstract}









\begin{keyword}



Overlapping community detection \sep Semi-supervised learning \sep Graph neural networks \sep Graph transformer \sep Pseudo-labels \sep Weak cliques

\end{keyword}

\end{frontmatter}



\section{Introduction}
Community structure is a fundamental characteristic of complex networks, widely present in various fields such as social, biological, and information networks
~\cite{Chen_2010,NICOLINI201728,10.1007/s11280-021-00966-4}. 
The community is the group of nodes that are more densely connected compared with the nodes outside the group~\cite{Fortunato_2016}.
Community detection (CD) aims to uncover the underlying community structure of networks, revealing how nodes form tightly connected groups. 
Numerous CD methods~\cite{liu2024communitydetectiongraphneural,Wu_2022} have been successfully applied to identify distinct, non-overlapping community structures, making significant contributions to the understanding of community structures.
However, real-world networks often contain overlapping community structures~\cite{yang2012structureoverlapscommunitiesnetworks}, where nodes belong to multiple groups.
As a result, these overlapping communities cannot be effectively captured by non-overlapping CD methods.

In the past decade, researchers have increasingly focused on overlapping community detection (OCD), leading to significant advancements in this field. 
Algorithm-based methods like CPM~\cite{Palla_2005}, COPRA~\cite{Gregory_2010} and NMF~\cite{6341758} are popular solutions to OCD problem. 
CPM detects communities by merging adjacent k-cliques that share common nodes, while COPRA assigns labels based on neighborhood consensus, adjusting the belonging coefficient to control the degree of community overlap.
Meanwhile, NMF-based methods~\cite{PhysRevE.83.066114,8970691} derive soft community memberships by factorizing the adjacency matrix into low-rank factor matrices.
However, these approaches rely solely on link information, overlooking valuable attributes and prior information, which often results in limited effectiveness and practicality for large-scale networks.

NOCD~\cite{shchur2019overlappingcommunitydetectiongraph} is the first method that combines graph neural networks (GNNs) with the Bernoulli-Poisson probabilistic model for the OCD problem, achieving improved detection performance. 
However, employing a two-layer graph convolutional network (GCN) inherently limits its capacity to capture long-range node relationships.
Based on NOCD, DynaResGCN~\cite{muttakin2023overlappingcommunitydetectionusing} addresses the limitations of shallow GCNs by introducing up to forty layers of deep residual GCN with dynamic dilated aggregation for the OCD problem, but this approach significantly increases computational overhead and the risk of
over-smoothing, especially in large-scale networks.
Furthermore, most existing OCD methods rely solely on link and attribute information, placing relatively little emphasis on prior information (e.g., known node-community labels).
SSGCAE~\cite{HE20221464} addresses this limitation by incorporating prior information into its learning process. It employs a convolutional autoencoder that integrates link and attribute information while introducing a semi-supervised loss to enhance detection accuracy.
However, it only utilizes prior information as an auxiliary loss term, limiting its ability to fully leverage this valuable external knowledge.
Driven by the limitations of existing OCD methods, we identify five key \textbf{motivations} in this work.

\textbf{Fully Utilizing Prior Information.} 
Various CD methods~\cite{LIU2017304,6985550} have shown that incorporating prior information is critical for detection accuracy. 
In contrast, most existing OCD methods overlook the utilization of prior information, thereby limiting their overall effectiveness.

\textbf{Making Full Use of Link Information.} 
Algorithmic methods, such as CPM, effectively predict community affiliations for certain nodes by leveraging link information. 
However, many existing learning-based methods rely solely on neighborhood aggregation within GNNs, leading to incomplete utilization of link information.

\textbf{Performance Instability.} 
In real-world applications, prior information is typically available for only a small subset of nodes, making semi-supervised models a more practical choice. 
Nevertheless, the incompleteness of prior information negatively affects the generalization ability of these models, leading to performance instability.

\textbf{Limitations of GNNs.}
Existing GNN-based OCD models are constrained by limited receptive fields~\cite{alon2021bottleneckgraphneuralnetworks,wu2022representing} and over-smoothing~\cite{zhao2020pairnorm,chen2019measuring}, lacking efficient solutions to address these issues. 
For instance, DynaResGCN~\cite{muttakin2023overlappingcommunitydetectionusing} attempts to expand the receptive field by stacking up to forty layers of GNN.
However, this strategy significantly increases computational overhead and the risk of over-smoothing, especially in large-scale networks.

\textbf{Lack of Semi-supervised Baselines.}
While previous studies~\cite{Wu_2022,DANESHFAR2024108215,HE20221464} have demonstrated that incorporating limited supervision can enhance model performance, most existing OCD methods remain unsupervised. 
To the best of our knowledge, SSGCAE is the only available learning-based semi-supervised OCD baseline. 
This shortage limits comprehensive benchmarking and further development in in the OCD field.

Based on the above motivations, we propose a Weak-clique-based Overlapping Community Detection method, namely WOCD. 
WOCD first assigns high-quality pseudo-labels to a subset of nodes by identifying weak cliques, a form of cohesive subgraph. 
After that, WOCD employs a novel semi-supervised framework with pseudo-labels to efficiently utilize prior information. 
Additionally, WOCD uses a single-layer Graph Transformer combined with GNN to improve performance while maintaining efficiency. 

The main contributions of our work are summarized as follows:

\begin{itemize}
\item We propose a novel semi-supervised framework that integrates link, attribute, and prior information for overlapping community detection.
By incorporating pseudo-labels, the framework adapts effectively to conditions with limited labeled data, making it well-suited for real-world applications.

\item We propose a pseudo-label initialization module based on weak cliques.
WOCD can generate high-quality pseudo-labels before the training phase by leveraging multi-faceted information within the graph, leading to enhanced generalization ability. 

\item We combine a lightweight Graph Transformer with GNN to expand the receptive field and alleviate over-smoothing issue. 
This combination significantly improves detection accuracy while maintaining efficiency.

\item We address the lack of semi-supervised baselines by effectively incorporating semi-supervised losses into multiple unsupervised OCD methods, which enables a fair comparison with WOCD.

\item We conduct extensive evaluations on eight datasets. 
Compared with six existing OCD baselines, WOCD shows a significant accuracy advantage. 
Additionally, ablation studies confirm the performance improvement brought by each component of WOCD. 

\end{itemize}


\section{Related Work}
\label{sec_related_works}


\subsection{Overlapping Community Detection}

Traditional overlapping community detection (OCD) methods are typically divided into node clustering and link clustering approaches~\cite{SHI2013394}. 
Node clustering detects communities by grouping nodes based on their structural features, focusing on the properties and relationships of individual nodes~\cite{Lancichinetti_2009,lee2010detectinghighlyoverlappingcommunity}. 
In contrast, link clustering first clusters the edges in a network and then maps the resulting link communities to node communities by grouping the nodes that are incident to the edges within each link community~\cite{Ahn_2010,SHI2013394}.
A key inspiration for WOCD is the Clique Percolation Method (CPM)~\cite{Palla_2005}, which identifies all k-cliques and merges adjacent cliques into the same community. 
Despite its effectiveness, finding all k-cliques in a graph is NP-hard, and merging them is also computationally expensive, limiting scalability in large-scale networks. 
Subsequent works~\cite{Kumpula_2008,8944086,6968420,8047969} aimed to reduce CPM’s complexity, with W-CPM~\cite{8047969} excelling in both efficiency and solution quality. 
It uses weak cliques instead of k-cliques and merges communities based on the overlap of nodes within weak cliques as well as the links connecting them.


In recent years, many studies have focused on attribute information and have employed graph neural networks (GNNs) to address the OCD Problem. 
NOCD~\cite{shchur2019overlappingcommunitydetectiongraph} is the first model to use GNNs combined with a Bernoulli-Poisson model for detecting overlapping communities in undirected graphs. 
Afterward, subsequent learning-based methods, such as DynaResGCN~\cite{muttakin2023overlappingcommunitydetectionusing} and UCoDe~\cite{moradan2023ucodeunifiedcommunitydetection}, mainly utilize link information and node features, achieving better detection performance compared to algorithm-based approaches. 
However, these learning-based models do not fully leverage prior information. 
SSGCAE~\cite{HE20221464} improved upon this by introducing a semi-supervised loss to train their model, resulting in enhanced detection accuracy. 
Inspired by SSGCAE, the starting point of our work is how to fully leverage link information, node features, and prior knowledge to achieve better detection accuracy.

\section{Preliminaries}
\label{sec_preliminaries}


\subsection{Problem Statement}

Given an undirected, unweighted graph \(\mathcal{G} = (\mathcal{V}, \mathcal{E})\), where \(\mathcal{V}\) is the set of nodes and \(\mathcal{E}\) represents the set of edges, let \(N = |\mathcal{V}|\) and \(M = |\mathcal{E}|\) denote the total number of nodes and edges, respectively. 
A node is denoted by \(v_i\), and an edge between nodes \(v_i\) and \(v_j\) is represented as \(e_{ij} = (v_i, v_j) \in \mathcal{E}\). 
The adjacency matrix \(\mathbf{A} \in \mathbb{R}^{N \times N}\) is defined as: \(\mathbf{A}_{ij} = 1\) if \(e_{ij} \in \mathcal{E}\), otherwise \(\mathbf{A}_{ij} = 0\). 
Additionally, \(\mathbf{X} \in \mathbb{R}^{N \times D}\) is the \(D\)-dimensional attribute matrix representing the node features, and \(\mathbf{C} \in \mathbb{R}^{N \times K}\) is the ground truth community affiliation matrix, where \(K\) denotes the number of communities. 
Notably, existing overlapping community detection methods can be divided into non-parametric and parametric approaches, distinguished by whether \(K\) is assumed in advance.
In this work, WOCD follows recent neural community solvers~\cite{shchur2019overlappingcommunitydetectiongraph,muttakin2023overlappingcommunitydetectionusing,moradan2023ucodeunifiedcommunitydetection,HE20221464} and is designed as a parametric method.

\textbf{Definition 1: Semi-supervised Parametric Overlapping Community Detection.} 
Given a graph \(\mathcal{G}\), the number of communities \(K\), a set of randomly sampled nodes \(\mathcal{V}_{\text{sampled}} \subset \mathcal{V}\), and their corresponding ground truth community affiliation matrix \(\mathbf{C}_{\text{sampled}}\), the objective is to assign each node to one or more of the \(K\) communities, with the option that some nodes may not belong to any community.

\textbf{Definition 2: Weak Cliques~\cite{8047969}.} 
Given a pair of connected nodes \(u\) and \(v\), the weak clique \(\mathcal{W}_{uv}\) is defined as:

\begin{equation}
\mathcal{W}_{uv} = \{u, v\} \cup \left(n_u \cap n_v\right)
\end{equation}
where \(n_u\) and \(n_v\) represent the sets of neighbors of nodes \(u\) and \(v\).

\section{Methodology}
\label{sec_methodology}

In this section, we introduce WOCD, a semi-supervised overlapping community detection model. 
WOCD aims to improve detection accuracy by effectively integrating link, attribute, and prior information.

\subsection{Overview of the Framework}


The framework of WOCD can be divided into four key stages: \textit{Pseudo-Label Initialization}, \textit{Initial Training}, \textit{Pseudo-Label Update}, and \textit{Refined Training}. 
In the first stage, WOCD identifies weak cliques based on the link information of the input graph and assigns pseudo-labels to a subset of nodes according to the community affiliations of labeled nodes within each weak clique. 
Once initialized, WOCD proceeds to the initial training with pseudo-labels, using a three-layer Graph Convolutional Network (GCN) combined with a one-layer Graph Transformer (GT) to predict the community affiliations for each node. 
The loss is computed based on the predicted community affiliation matrix, true labels, and pseudo-labels. 
After completing the initial training, WOCD uses the best model parameters to generate a new set of pseudo-labels. 
In the final stage, WOCD performs refined training with updated pseudo-labels, which produces the final predicted community affiliation matrix, \( \mathbf{C}_{\text{final}} \). 
Further details of each stage are provided in the following subsections.

\begin{figure*}[t]  
  \centering  
  \includegraphics[width=\textwidth]{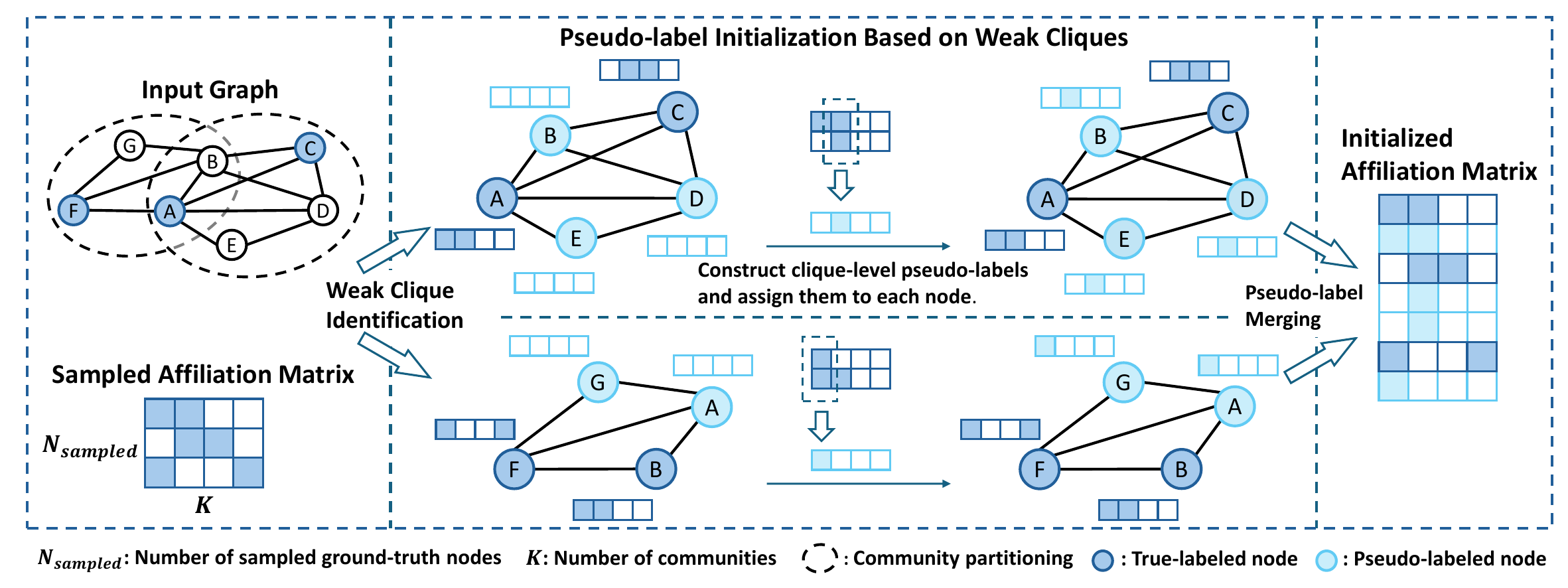}  
  \caption{The process of pseudo-label initialization based on weak cliques. Given an input graph and a random sampled affiliation matrix, WOCD first identifies weak cliques based on link information. Then, it constructs clique-level pseudo-labels by aggregating the ground-truth labels of sampled nodes in each weak clique. These pseudo-labels are subsequently assigned to all nodes in the corresponding weak cliques. After that, WOCD merges the pseudo-labels of each node to obtain the initialized overlapping community affiliation matrix.}
  \label{fig_WOCD_Weak_clique_identification_Framework}  
\end{figure*}

\subsection{Pseudo-Label Initialization}

To address the performance instability in current semi-supervised learning models caused by the insufficient true labeled nodes, we first initialize pseudo-labels based on structure information before training. 
Considering the close relationship between community assignment and graph cohesiveness, we choose to use weak cliques to initialize pseudo-labels, allowing for community overlap while also ensuring computational efficiency. 
Specifically, we assume that all nodes within a given weak clique share membership in at least one common community. 
Since weak cliques allow overlap, nodes may belong to multiple weak cliques, each with potentially different community affiliations. 
By merging these affiliations, we generate high-quality pseudo-labels for each node, effectively leveraging link information.
The pseudo-label initialization process consists of two main steps: \textit{Weak Clique Identification} and \textit{Pseudo-Label Construction}. The overall process of the pseudo-label initialization can be found in Figure~\ref{fig_WOCD_Weak_clique_identification_Framework}.

\begin{algorithm}[ht]
\caption{Weak Clique Identification}
\label{alg_identify_weak_cliques}
\begin{algorithmic}[1]
\REQUIRE Graph $G = (V, E)$

\ENSURE Weak clique set $\mathcal{W}$

\STATE $P \gets \varnothing$, $\mathcal{W} \gets \varnothing$

\FOR{each node $i \in V$}
    \STATE $m_i \gets$ number of edges between neighbors of node $i$
    \STATE $d_i \gets$ degree of node $i$
    \STATE $P_i \gets \frac{m_i + d_i}{d_i + 1}$ 
\ENDFOR

\WHILE{$P \neq \varnothing$}
    \STATE $u \gets \text{argmax}(P)$
    \STATE $n_u \gets$ set of neighbors of node $u$
    \STATE $SI_u \gets \varnothing$
    \FOR{$v \in n_u$}
        \STATE $n_v \gets$ set of neighbors of node $v$
        \STATE $SI_{uv} \gets \frac{|n_u \cap n_v|}{\sqrt{|n_u| \times |n_v|}}$ 
    \ENDFOR
    \STATE $v \gets \text{argmax}(SI_u)$
    \STATE $\mathcal{W}_{uv} \gets \{u, v\} \cup (n_u \cap n_v)$ 
    \STATE $\mathcal{W} \gets \mathcal{W} \cup \mathcal{W}_{uv}$
    \STATE Remove $u$ and $v$ from $P$
\ENDWHILE
\RETURN $\mathcal{W}$
\end{algorithmic}
\end{algorithm}

\textbf{Weak Clique Identification.}
Weak clique identification starts by selecting a starting node and its most similar neighbor.
To maximize the likelihood that all nodes in a resulting weak clique belong to the same community, it is crucial to carefully select this pair of adjacent nodes, particularly the starting node. 
To guide this selection, the priority of node \(u\) is defined as:

\begin{equation}
P_u = \frac{m_u + d_u}{d_u + 1},
\end{equation}
where \(m_u\) represents the number of links between the neighbors of node \(u\), and \(d_u\) denotes the degree of node \(u\). 
Next, we use the Salton Index~\cite{10.5555/576628} to find the adjacent node that has the greatest similarity to node \(u\). 
The Salton Index between nodes \(u\) and \(v\) is given by:
\begin{equation}
SI_{uv} = \frac{|n_u \cap n_v|}{\sqrt{|n_u| \times |n_v|}}
\end{equation}
where \(n_u\) and \(n_v\) represent the sets of neighbors of nodes \(u\) and \(v\), \(| \cdot |\) denotes the number of nodes in the set. 
A higher Salton Index indicates that two nodes share more common neighbors, increasing the likelihood that these nodes and their shared neighbors belong to the same community. 

Once the most similar node \(v\) for node \(u\) is selected, the nodes connected to both \(u\) and \(v\) are added to the weak clique \(\mathcal{W}_{uv}\).
Please note that once selected, \(u\) and \(v\) cannot be used as starting points but still can be selected in other weak cliques.
WOCD then iterates over the remaining nodes, identifying the highest priority node and its most similar neighbor to form subsequent weak cliques. 
This process continues until all nodes have been used as starting nodes for weak clique construction.
The complete procedure for weak clique identification is presented in Algorithm~\ref{alg_identify_weak_cliques}.
In addition, we present a simple example of this process in~\ref{app_architectural_details}.

\begin{algorithm}[ht]
\caption{Pseudo-Label Construction}
\label{alg_construct_pseudo}
\begin{algorithmic}[1]
\REQUIRE Weak clique set $\mathcal{W}$, sampled node set $\mathcal{V}_{\text{sampled}}$, sampled true community affiliation matrix $\mathbf{C}_{\text{sampled}}$, retained number of communities $r_{c}$
\ENSURE Pseudo-labels $\mathbf{C}_{\text{pseudo}}$

\STATE Initialize $\mathbf{C}_{\text{pseudo}}$ as a zero matrix

\FOR{each weak clique $\mathcal{W}_{i}$ in $\mathcal{W}$}
    \STATE $\mathbf{C}_{\text{clique}} \gets \mathbf{0}$ 
    \FOR{each node $u \in \mathcal{W}_{i}$}
        \IF{$u \in \mathcal{V}_{\text{sampled}}$}
            \STATE $\mathbf{C}_{\text{clique}} \gets \mathbf{C}_{\text{clique}} + \mathbf{C}_{\text{sampled}}[u]$  
        \ENDIF
    \ENDFOR
    \FOR{each community $j$ in $\mathbf{C}_{\text{clique}}$}
        \IF{$j \in$ top $r_{c}$ largest communities}
            \STATE $\mathbf{C}_{\text{clique}}[j] \gets 1$
        \ELSE
            \STATE $\mathbf{C}_{\text{clique}}[j] \gets 0$
        \ENDIF
    \ENDFOR

    \FOR{each node $v \in \mathcal{W}_{i}$}
        \STATE $\mathbf{C}_{\text{pseudo}}[v] \gets \mathbf{C}_{\text{pseudo}}[v] + \mathbf{C}_{\text{clique}}$
    \ENDFOR
\ENDFOR
\STATE Set all elements in $\mathbf{C}_{\text{pseudo}}$ greater than 0 to 1
\RETURN $\mathbf{C}_{\text{pseudo}}$
\end{algorithmic}
\end{algorithm}

\textbf{Pseudo-Label Construction.} 
After identifying weak cliques, a key step is to use the labeled node set \(\mathcal{V}_{\text{sampled}}\) to assign high-quality pseudo-labels to certain nodes based on cohesive graph structure. 
This process first determines the most likely community label for each weak clique based on the labeled nodes within \(\mathcal{V}_{\text{sampled}}\). 
These labels serve as the label for each node within the weak clique. 
Notably, the same node can appear in multiple weak cliques. 
For instance, if a node \(u\) belongs to three weak cliques which are assigned to different communities, the pseudo-label for node \(u\) will include all assigned communities. 
Algorithm \ref{alg_construct_pseudo} illustrates the detailed process of constructing pseudo-labels.


\begin{figure*}[t]  
  \centering  
  \includegraphics[width=\textwidth]{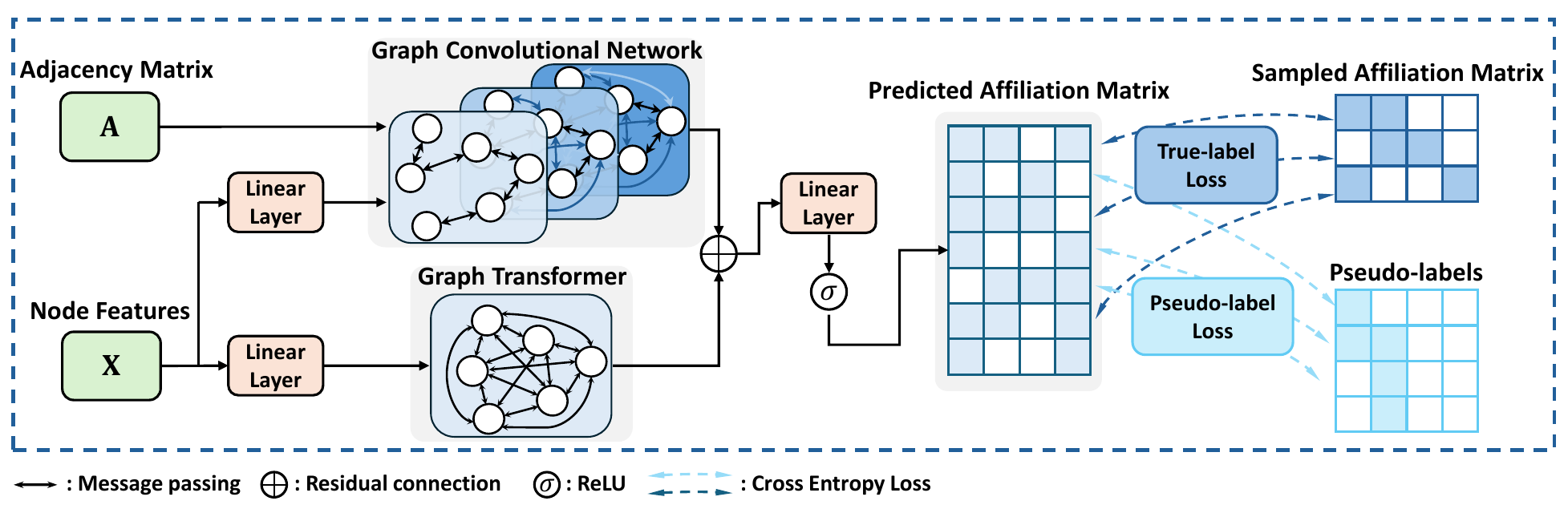}  
  \caption{An overview of the training framework.}
  \label{fig_WOCD_Network_Framework}  
\end{figure*}

\subsection{Initial Training}

After initializing pseudo-labels, we apply a learning framework that combines a three-layer Graph Convolutional Network (GCN) with a single-layer Graph Transformer (GT). 
The overall network architecture is illustrated in Figure~\ref{fig_WOCD_Network_Framework}.
By combining GCN's local feature aggregation with GT's ability to capture long-range dependencies, WOCD effectively leverages link and attribute information to predict the community affiliation matrix, \(\mathbf{C}_{\text{pred}}\), as follows:

\begin{equation}
\begin{aligned}
&\mathbf{Z}_{\text{GCN}} = \text{GCN}(\mathbf{A}, \mathbf{X}), \quad \mathbf{Z}_{\text{GT}} = \text{GT}(\mathbf{X}), \\
&\mathbf{C}_{\text{pred}} = f_\mathbf{Z}(\alpha \mathbf{Z}_{\text{GCN}} + \beta \mathbf{Z}_{\text{GT}})
\end{aligned}
\end{equation}
In this equation, \(\alpha\) and \(\beta\) are hyperparameters controlling the contributions of the GCN and GT embeddings, respectively. 
\(f_\mathbf{Z}\) is a linear layer that maps the aggregated node embeddings to \(K\) distinct communities. The combination of \(\mathbf{Z}_{\text{GCN}}\) and \(\mathbf{Z}_{\text{GT}}\) enables the model to effectively integrate both GCN and GT representations.

\textbf{Graph Convolutional Network~\cite{kipf2017graphconvolutional}.} A single-layer Graph Convolutional Network (GCN) is formulated as follows:

\begin{equation}
\mathbf{Z}^{(l+1)} = \sigma \left( \tilde{\mathbf{D}}^{-\frac{1}{2}} \tilde{\mathbf{A}} \tilde{\mathbf{D}}^{-\frac{1}{2}} \mathbf{Z}^{(l)} \mathbf{\Theta}^{(l)} \right)
\end{equation}
where \(\tilde{\mathbf{A}} = \mathbf{A} + \mathbf{I}\) is the adjacency matrix with added self-loops, and \(\tilde{\mathbf{D}}\) is the degree matrix of \(\tilde{\mathbf{A}}\). 
In this equation, \(\mathbf{Z}^{(l)}\) denotes the input node embeddings at layer \(l\), \(\mathbf{\Theta}^{(l)}\) represents the learnable weight matrix at layer \(l\), and \(\sigma\) is the non-linear activation function (e.g., ReLU), introducing non-linearity.

\textbf{Graph Transformer.} 
As an emerging graph encoder, the Graph Transformer demonstrates strong performance on small graphs. However, as the number of nodes increases, its computational overhead also grows, potentially limiting its generalization capability on large-scale networks. 
Moreover, applying Graph Transformers often requires stacking multiple attention layers, which increases network complexity. 
Inspired by SGFormer~\cite{wu2024sgformersimplifyingempoweringtransformers}, we utilize a single-layer, single-head Graph Transformer, where \(\mathbf{Q}\), \(\mathbf{K}\), and \(\mathbf{V}\) are defined as follows:

\begin{equation}
\begin{aligned}
&\mathbf{Q} = f_{\mathbf{Q}}(\mathbf{Z}^{(0)}), \quad \tilde{\mathbf{Q}} = \frac{\mathbf{Q}}{\|\mathbf{Q}\|_{\mathcal{F}}}, \\
&\mathbf{K} = f_{\mathbf{K}}(\mathbf{Z}^{(0)}), \quad \tilde{\mathbf{K}} = \frac{\mathbf{K}}{\|\mathbf{K}\|_{\mathcal{F}}}, \\
&\mathbf{V} = f_{\mathbf{V}}(\mathbf{Z}^{(0)}),
\end{aligned}
\end{equation}
where \(f_\mathbf{Q}\), \(f_\mathbf{K}\), and \(f_\mathbf{V}\) are linear layers. 
\(\|\mathbf{Q}\|_{\mathcal{F}}\) denotes the Frobenius norm of \(\mathbf{Q}\). 
The resulting linear attention function is defined as follows:

\begin{equation}
\begin{aligned}
&\mathbf{D} = \text{diag} \left( \mathbf{1} + \frac{1}{N} \tilde{\mathbf{Q}} \left( \tilde{\mathbf{K}}^\top \mathbf{1} \right) \right), \\
&\mathbf{Z}_{\text{GT}} = \gamma \mathbf{D}^{-1} \left[ \mathbf{V} + \frac{1}{N} \tilde{\mathbf{Q}} \left( \tilde{\mathbf{K}}^\top \mathbf{V} \right) \right] + (1 - \gamma) \mathbf{Z}^{(0)}
\end{aligned}
\end{equation}
where \(\mathbf{1}\) is an \(N\)-dimensional all-ones column vector, and the \text{diag} operation converts the \(N\)-dimensional column vector into an \(N \times N\) diagonal matrix. 
\(\gamma\) is the hyperparameter for the residual connection, which helps prevent over-smoothing and maintains the initial input features.

\textbf{Loss Function.} 
The learning objective in WOCD is to minimize the binary cross-entropy (BCE) loss for both the sampled nodes with true labels and the remaining nodes with pseudo-labels. 
Specifically, the loss function is defined as:

\begin{equation}
\label{equ_loss_function}
\mathcal{L} = \lambda_1 \cdot \text{BCE}(\mathbf{C}_{\text{pred}}, \mathbf{C}_{\text{sampled}}) + \lambda_2 \cdot \text{BCE}(\mathbf{C}_{\text{pred}}, \mathbf{C}_{\text{pseudo}})
\end{equation}
where \(\lambda_1\) and \(\lambda_2\) are hyperparameters that balance the contributions of the true-labeled and pseudo-labeled nodes in the loss function.




\subsection{Refined Training}
Since the pseudo-labels generated by weak cliques do not cover the majority of nodes, we retain the last-performing model parameters from the initial training and use them to generate new pseudo-labels. 
To improve the quality of pseudo-labels, we introduce a pseudo-label confidence threshold \(\tau\), which filters out low-quality pseudo-labels and retains only the most reliable ones. 

After updating the pseudo-labels, WOCD proceeds to refined training. 
The same network structure and loss function as in the initial training are used here. 
The key difference between the two rounds of training is that the pseudo-labels in refined training cover a larger proportion of nodes, thereby enhancing the model’s generalization capability.

\subsection{Time Complexity Analysis}
The time complexity of WOCD consists of two parts: weak clique construction and model training.

Unlike k-cliques, the construction of weak cliques starts with a node and its most similar neighbor, with each node serving as a starting point only once.
For a node \(u\) serving as the starting point of a weak clique, the time complexity of finding its most similar node is \(O(d_u)\). 
The overall time complexity for constructing weak cliques is proportional to the sum of the degrees of the nodes, resulting in a total time complexity of \(O(M)\).


In the training phase, WOCD employs a three-layer GCN and a single-layer Graph Transformer. The time complexity of a single-layer GCN is:
\begin{equation}
O \left( M \cdot h + N \cdot h^2 \right) 
\end{equation}
where \(h\) represents the hidden dimension. The time complexity of a one-layer GT is: 
\begin{equation}
O\left( N \cdot h^2 \right).
\end{equation}

\section{Experiment}
\label{sec_experiment}

In this section, we present a comparative analysis of the WOCD model against six baseline methods across eight real-world attributed graphs. 
Additionally, we conduct ablation studies to demonstrate that the key factors contributing to WOCD's effectiveness are the use of pseudo-labels, weak cliques, and the Graph Transformer. 

\noindent
\begin{minipage}[t]{0.44\textwidth}
\begin{table}[H]
\centering
\small
\renewcommand{\arraystretch}{1.2}
\setlength{\tabcolsep}{5pt}
\resizebox{\textwidth}{!}{
\begin{tabular}{lrrrrr}
\toprule
\textbf{Dataset} & \(N\) & \(M\) & \(D\) & \(K\) \\
\midrule
FB0     & 347   & 2.5K    & 30   & 24  \\
FB107   & 1.0K  & 26.7K   & 11   & 9   \\
FB1684  & 792   & 14.0K   & 15   & 17  \\
FB1912  & 755   & 30.0K   & 29   & 46  \\
\midrule
Chem     & 35.4K & 157.4K  & 4.9K & 14  \\
CS       & 22.0K & 96.8K   & 7.8K & 18  \\
Eng      & 14.9K & 49.3K   & 4.8K & 16  \\
Med      & 63.3K & 810.3K  & 5.5K & 17  \\
\bottomrule
\end{tabular}
}
\caption{Dataset statistics.}
\label{tab_dataset_statistics}
\end{table}
\end{minipage}
\hfill
\begin{minipage}[t]{0.53\textwidth}
\begin{table}[H]
\centering
\small
\renewcommand{\arraystretch}{1.2}
\setlength{\tabcolsep}{4pt}
\resizebox{\textwidth}{!}{
\begin{tabular}{llcccc}
\toprule
\textbf{Type} & \textbf{Algorithm} & \(\mathbf{A}\) & \(\mathbf{X}\) & \(K\) & \(\mathbf{C}_{\text{sampled}}\) \\
\midrule
\multirow{2}{*}{\makecell{Non-parammetric\\OCD Methods}} 
& W-CPM~\cite{8047969}            & \cmark & \xmark & \xmark & \xmark \\
& GREESE~\cite{10.1007/s00607-021-00948-4}           & \cmark & \xmark & \xmark & \xmark \\
\midrule
\multirow{3}{*}{\makecell{Unsupervised\\OCD Methods}} 
& NOCD~\cite{shchur2019overlappingcommunitydetectiongraph}             & \cmark & \cmark & \cmark & \xmark \\
& DynaResGCN~\cite{muttakin2023overlappingcommunitydetectionusing}       & \cmark & \cmark & \cmark & \xmark \\
& UCoDe~\cite{moradan2023ucodeunifiedcommunitydetection}            & \cmark & \cmark & \cmark & \xmark \\
\midrule
\multirow{5}{*}{\makecell{Semi-supervised\\OCD Methods}} 
& NOCD + SSL        & \cmark & \cmark & \cmark & \cmark \\
& DynaResGCN + SSL  & \cmark & \cmark & \cmark & \cmark \\
& UCoDe + SSL       & \cmark & \cmark & \cmark & \cmark \\
& SSGCAE~\cite{HE20221464}           & \cmark & \cmark & \cmark & \cmark \\
& WOCD w/o Pseudo           & \cmark & \cmark & \cmark & \cmark \\
& \textbf{WOCD}    & \cmark & \cmark & \cmark & \cmark \\
\bottomrule
\end{tabular}
}
\caption{Information Required by Baselines.}
\label{tab_input_information}
\end{table}
\end{minipage}

\subsection{Experimental Setting}
\textbf{Datasets.} We select eight real-world overlapping datasets for our experiments, with their specific statistics listed in Table \ref{tab_dataset_statistics}. 
The Facebook datasets~\cite{10.1145/2556612} are ego-nets derived from Facebook. 
We also employ four large-scale Mag datasets used in the experiments of NOCD~\cite{shchur2019overlappingcommunitydetectiongraph}.

\textbf{Baselines.}  
Although recent unsupervised and semi-supervised methods~\cite{Wu_2022,9722614,wang2024efficientunsupervisedcommunitysearch,wang2024neuralattributedcommunitysearch} have demonstrated significant effectiveness in identifying community structures, we did not include them in our experiments. This is because these methods are designed for non-overlapping communities and are not well-suited for the OCD problem addressed in our work.

Since the only available state-of-the-art semi-supervised method for the OCD problem is SSGCAE, the lack of diverse semi-supervised baselines limits a comprehensive evaluation of WOCD.
To address this gap, we extend three existing unsupervised methods by incorporating semi-supervised learning (SSL), namely NOCD + SSL, DynaResGCN + SSL, and UCoDe + SSL.
The way we introduce supervision is aligned with SSGCAE to ensure a fair comparison across models.
To verify the effectiveness and stability of our weak clique-based pseudo-labeling strategy, we also build WOCD w/o Pseudo, a variant of WOCD that shares the same network architecture but is trained solely using ground truth labels without leveraging any pseudo-labels.

The categorization and input requirements of all baseline models are summarized in Table~\ref{tab_input_information}.
Due to space limitations, details of all baseline models and the integration of semi-supervised losses are provided in~\ref{app_baselines}.

\begin{table}[H]
\centering
\small
\setlength{\tabcolsep}{4pt}
\renewcommand{\arraystretch}{1.2}

\resizebox{\textwidth}{!}{%
\begin{tabular}{l|cc|cc|cc|cc}
\toprule
\textbf{Algorithm} 
& \multicolumn{2}{c|}{\textbf{FB0}} 
& \multicolumn{2}{c|}{\textbf{FB107}} 
& \multicolumn{2}{c|}{\textbf{FB1684}} 
& \multicolumn{2}{c}{\textbf{FB1912}}\\
& ONMI (\%) ↑ & Time ↓ & ONMI (\%) ↑ & Time ↓ & ONMI (\%) ↑ & Time ↓ & ONMI (\%) ↑ & Time ↓ \\
\midrule
W-CPM & 5.8 & 3.32s & 6.9 & 4.34s & 22.8 & 3.68s & 23.4 & 4.91s \\
GREESE & 6.4 & 3.61s & 8.7 & 6.98s & 40.8 & 5.14s & 28.6 & 6.26s \\
\midrule
NOCD & 13.9 ± 2.1 & 0.11s & 12.5 ± 0.5 & 0.12s & 26.1 ± 1.3 & 0.12s & 35.6 ± 1.3 & 0.21s \\
DynaResGCN & 12.5 ± 0.5 & 0.24s & 12.6 ± 1.5 & 0.81s & 44.3 ± 2.6 & 0.78s & 40.1 ± 1.7 & 0.39s \\
UCoDe & 10.8 ± 1.2 & 0.01s & 13.6 ± 2.4 & 0.03s & 38.4 ± 1.3 & 0.02s & 37.3 ± 1.6 & 0.02s \\
\midrule
NOCD + SSL & \underline{15.6 ± 1.2} & \underline{0.11s} & 13.9 ± 1.5 & 0.12s & 37.2 ± 4.1 & 0.12s & 36.1 ± 3.6 & 0.21s \\
DynaResGCN + SSL & 13.8 ± 1.1 & 0.24s & 15.3 ± 1.3 & 0.81s & \underline{49.5 ± 3.0} & \underline{0.78s} & 37.5 ± 2.8 & \underline{0.39s} \\
UCoDe + SSL & 11.7 ± 1.0 & 0.01s & 14.1 ± 3.0 & 0.03s & 38.1 ± 1.1 & 0.02s & 37.9 ± 2.3 & 0.02s \\
SSGCAE & 14.4 ± 1.6 & 0.04s & \textbf{23.7 ± 1.8} & \textbf{0.02s} & 42.6 ± 3.9 & 0.03s & 34.3 ± 2.2 & 0.13s\\
WOCD w/o Pseudo & 14.5 ± 2.5 & 0.04s & 17.5 ± 1.8 & 0.02s & 36.4 ± 4.5 & 0.03s & \underline{41.1 ± 2.6} & \underline{0.12s}\\
WOCD & \textbf{19.9 ± 1.9} & \textbf{0.07s} & \underline{19.8 ± 1.6} & \underline{0.07s} & \textbf{59.3 ± 3.4} & \textbf{0.07s} & \textbf{42.1 ± 1.7} & \textbf{0.16s} \\
\bottomrule
\end{tabular}
}
\caption{Results on four Facebook Datasets}
\label{tab_fb_results}
\end{table}

\vspace{-12pt}

\begin{table}[H]
\centering
\small
\setlength{\tabcolsep}{4pt}
\renewcommand{\arraystretch}{1.2}

\resizebox{\textwidth}{!}{%
\begin{tabular}{l|cc|cc|cc|cc}
\toprule
\textbf{Algorithm} 
& \multicolumn{2}{c|}{\textbf{Chemistry}} 
& \multicolumn{2}{c|}{\textbf{Computer Science}} 
& \multicolumn{2}{c|}{\textbf{Engineering}} 
& \multicolumn{2}{c}{\textbf{Medicine}}\\
& ONMI (\%) ↑ & Time ↓ & ONMI (\%) ↑ & Time ↓ & ONMI (\%) ↑ & Time ↓ & ONMI (\%) ↑ & Time ↓ \\
\midrule
W-CPM & 0.0 & 11.18m & 0.0 & 5.10m & 0.0 & 2.30m & 0.0 & 24.86m \\
GREESE & 0.0 & 18.31m & 0.0 & 9.51m & 0.0 & 3.43m & 0.0 & 38.58m \\
\midrule
NOCD & 45.3 ± 2.3 & 2.10s & 50.2 ± 2.0 & 2.27s & 39.1 ± 4.5 & 1.22s & 37.8 ± 2.8 & 5.70s \\
DynaResGCN & 30.3 ± 2.9 & 12.58s & 37.4 ± 1.7 & 6.78s & 37.3 ± 1.2 & 8.73s & 40.0 ± 2.1 & 29.23s \\
UCoDe & 33.7 ± 2.1 & 23.72s & 42.5 ± 1.7 & 9.20s & 33.2 ± 0.9 & 4.31s & 32.7 ± 3.7 & 76.75s \\
\midrule
NOCD + SSL & \underline{54.1 ± 0.4} & \underline{2.10s} & 58.1 ± 3.2 & 2.27s & 60.9 ± 2.1 & 1.22s & 55.2 ± 1.9 & 5.70s \\
DynaResGCN + SSL & 38.7 ± 1.9 & 12.58s & 50.0 ± 1.3 & 6.78s & 49.0 ± 1.3 & 8.73s & 49.1 ± 1.8 & 29.23s \\
UCoDe + SSL & 50.7 ± 3.0 & 23.72s & 55.7 ± 2.9 & 9.20s & 55.1 ± 2.9 & 4.31s & 51.6 ± 2.4 & 76.75s \\
SSGCAE & 51.2 ± 0.6 & 1.03s & 53.5 ± 0.9 & 0.56s & \underline{65.1 ± 0.7} & \underline{0.33s} & OOM & OOM \\
WOCD w/o Pseudo & 53.8 ± 3.0 & 4.93s & \underline{58.4 ± 2.6} & \underline{3.92s} & 60.7 ± 4.1 & 2.21s & \underline{57.0 ± 2.2} & \underline{16.03s} \\
WOCD & \textbf{59.2 ± 1.4} & \textbf{5.72s} & \textbf{61.9 ± 1.5} & \textbf{4.91s} & \textbf{65.8 ± 1.1} & \textbf{2.35s} & \textbf{59.3 ± 1.5} & \textbf{21.33s} \\
\bottomrule
\end{tabular}
}
\caption{Results on four large-scale MAG Datasets}
\label{tab_MAG_results}
\end{table}

\textbf{Parameter Settings.} 
We describe the parameter settings of WOCD in this paragraph. Specifically, we employ a three-layer GCN and a single-layer, single-head GT. 
The dimension of the initial features is set to 256, and the output dimension is determined by the number of communities \(K\). 
Adam~\cite{kingma2017adammethodstochasticoptimization} is selected as the optimizer for WOCD, with the learning rate set to  $1\times 10^{-3}$. 
The loss balance coefficients \(\lambda_1\) and \(\lambda_2\) in Equation \ref{equ_loss_function} are tuned within the range from $1$ to $5$, and the training epochs range from $150$ to $1000$ for different datasets. 
The sampling strategy is designed to select an equal number of nodes from each community, which aligns with real-world applications.
To ensure fair comparison among all semi-supervised models, we control the proportion of labeled nodes via a sampling ratio denoted by \(\rho\).
All experiments were conducted using an Intel(R) Xeon(R) Gold 6348 CPU and an NVIDIA A800 GPU with 80GB memory.

\textbf{Evaluation Metrics.} 
We evaluate the performance of WOCD against baseline models in terms of both accuracy and efficiency. Accuracy is measured using Overlapping Normalized Mutual Information (ONMI)~\cite{mcdaid2013normalizedmutualinformationevaluate}, which is consistent with most OCD methods. 
For efficiency, we report the average running time per epoch for each model. 
The experimental results are presented in the following subsections.

\subsection{Performance Analysis}


The experimental results are represented in Table~\ref{tab_fb_results} and Table~\ref{tab_MAG_results}. 
After integrating semi-supervised learning, all three previously unsupervised methods exhibit improved performance across all datasets, highlighting the importance of leveraging prior information for accurate community detection.

Under the same prior settings, WOCD consistently outperforms the five semi-supervised baselines in terms of both accuracy and stability, while maintaining competitive efficiency. 
In addition, compared with its variant, WOCD consistently achieves higher accuracy and stability across all datasets, owing to its ability to expand supervision to a broader range of nodes through weak clique-based pseudo-label generation. 
This extended supervision enables WOCD to better generalize across varying graph structures and community distributions.
Overall, these results demonstrate that WOCD effectively combines structural, attribute, and limited prior information to achieve robust and accurate overlapping community detection.

\begin{figure*}[h!]
    \centering
    \begin{subfigure}[b]{0.49\textwidth}
        \includegraphics[width=\textwidth]{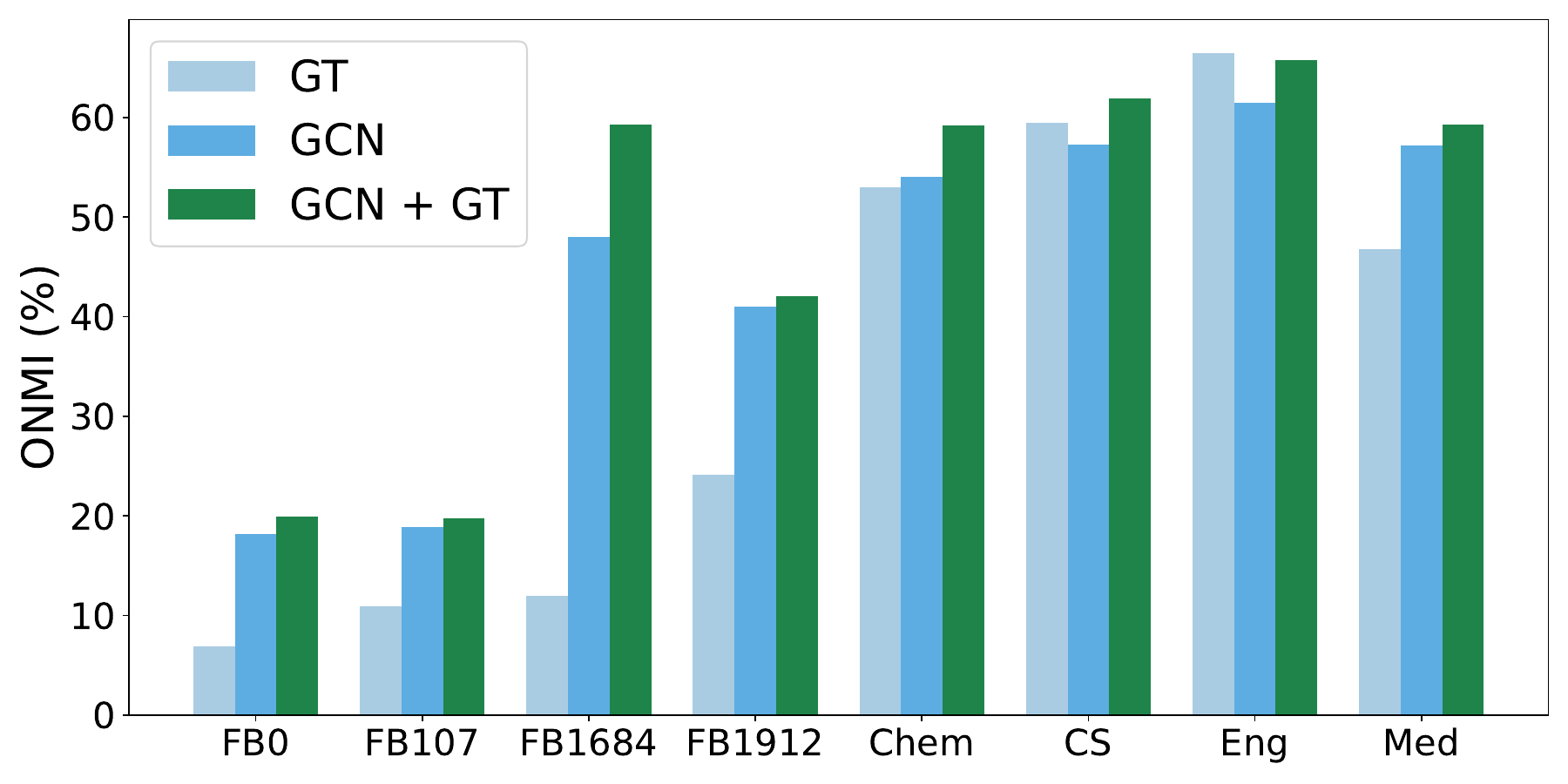}
        \caption{Impact of GT and GCN on accuracy.}
        \label{fig_GCN_GT_accuracy}
    \end{subfigure}
    \hfill
    \begin{subfigure}[b]{0.49\textwidth}
        \includegraphics[width=\textwidth]{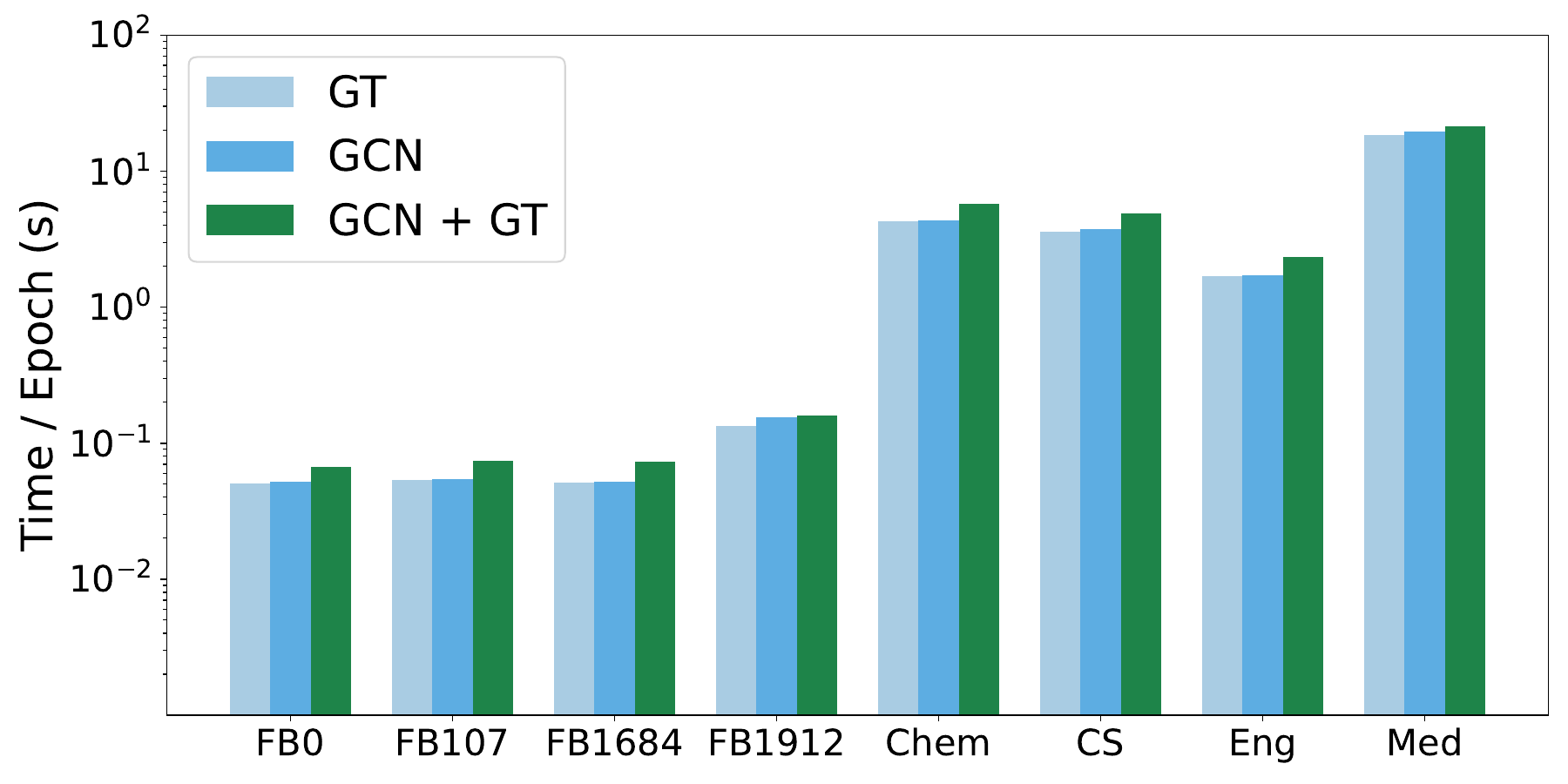}
        \caption{Impact of GT and GCN on efficiency.}
        \label{fig_GCN_GT_efficiency}
    \end{subfigure}

    \vskip\baselineskip
    
    \begin{subfigure}[b]{0.98\textwidth}
        \centering
        \includegraphics[width=\textwidth]{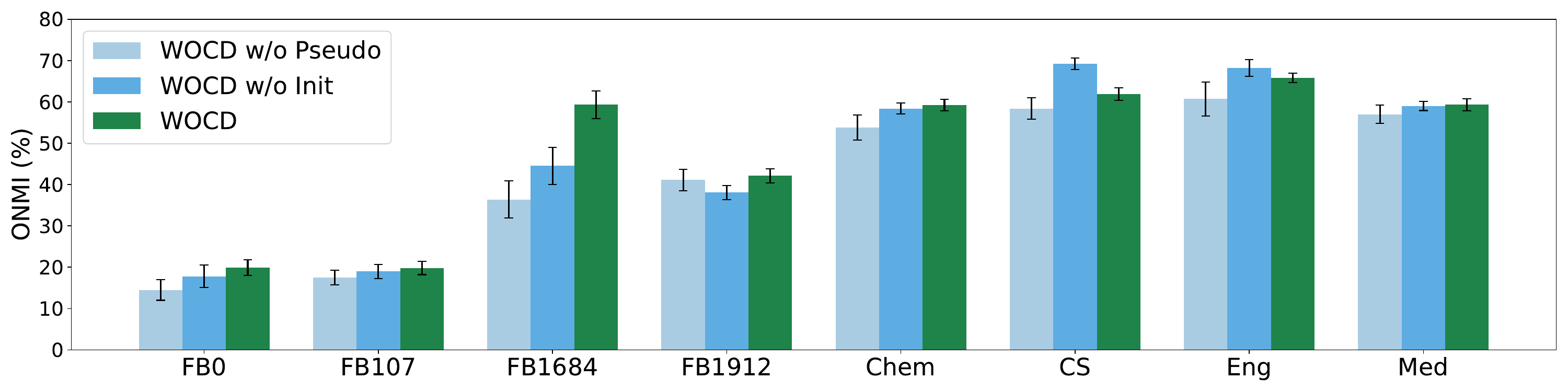}
        \caption{Impact of using and initializing pseudo-labels on performance.}
        \label{fig_pseudo_ablation}
    \end{subfigure}

    \caption{Ablation study of the WOCD model.}
\end{figure*}

\subsection{Ablation Experiments}
In this subsection, we conduct a series of ablation experiments to demonstrate that the key factors contributing to WOCD's high detection accuracy are the introduction of Graph Transformer, the use of pseudo-labels, and the initialization of pseudo-labels using weak cliques.
In addition, we perform an additional experiment to verify the effectiveness of introducing the refined training phase in improving both accuracy and stability of the model.

\textbf{Exp-1: Graph Transformer.} 
In our work, we incorporate a single-layer Graph Transformer (GT) to enhance the model's receptive field and adapt to more complex overlapping community structures. 
We conducted experiments using both a three-layer GCN and a single-layer GT. 
The performance and efficiency results are presented in Figures \ref{fig_GCN_GT_accuracy} and \ref{fig_GCN_GT_efficiency}, respectively. 
The ablation study results demonstrate that by combining a single-layer GT with GCN, WOCD achieves the highest accuracy across all eight datasets. 
Despite the addition of a single-layer GT, WOCD maintains a running time similar to that of a three-layer GCN alone.


\textbf{Exp-2: Necessity of Using Pseudo-labels.} 
To verify the effect of using pseudo-labels, we compare WOCD with its variant WOCD w/o Pseudo.
As shown in Table~\ref{fig_pseudo_ablation}, incorporating pseudo-labels enables WOCD to achieve higher ONMI scores and the lowest standard deviation across all datasets, demonstrating the effectiveness and stability of our pseudo-labeling strategy.

\textbf{Exp-3: Necessity of Pseudo-Label Initialization.} 
One of the key contributions of our work is the use of weak cliques to initialize pseudo-labels before training. 
To assess its impact, we remove the pseudo-label initialization process and refer to this variant as WOCD w/o Init.
The results in Table~\ref{fig_pseudo_ablation} indicate that the model with pseudo-label initialization achieves higher performance on 6 out of 8 datasets, demonstrating the effectiveness of leveraging weak cliques.

\begin{table}[H]
\centering
\resizebox{\textwidth}{!}{%
\begin{tabular}{l|cc|cc|cc|cc}
\toprule
\textbf{Round} 
& \multicolumn{2}{c|}{\textbf{FB0}} 
& \multicolumn{2}{c|}{\textbf{FB107}} 
& \multicolumn{2}{c|}{\textbf{FB1684}} 
& \multicolumn{2}{c}{\textbf{FB1912}} \\
& \(N_{\text{pseudo}}\) & ONMI (\%) 
& \(N_{\text{pseudo}}\) & ONMI (\%) 
& \(N_{\text{pseudo}}\) & ONMI (\%) 
& \(N_{\text{pseudo}}\) & ONMI (\%) \\
\midrule
Initial Training  & 126 & 19.3 ± 2.9 & 603 & 19.7 ± 3.0 & 454 & \textbf{59.3 ± 3.4} & 500 & 41.4 ± 2.7 \\
Refined Training & 129 & \textbf{19.9 ± 1.9} & 409 & \textbf{19.8 ± 1.6} & 539 & 57.1 ± 4.7 & 473 & \textbf{42.1 ± 1.7} \\
\bottomrule
\end{tabular}%
}
\caption{Comparison results between first and second rounds on Facebook datasets.}
\label{tab_pseudo_generation_FB}
\end{table}

\begin{table}[H]
\centering
\resizebox{\textwidth}{!}{%
\begin{tabular}{l|cc|cc|cc|cc}
\toprule
\textbf{Round} 
& \multicolumn{2}{c|}{\textbf{CHEM}} 
& \multicolumn{2}{c|}{\textbf{CS}} 
& \multicolumn{2}{c|}{\textbf{ENG}} 
& \multicolumn{2}{c}{\textbf{MAD}} \\
& \(N_{\text{pseudo}}\) & ONMI (\%) 
& \(N_{\text{pseudo}}\) & ONMI (\%) 
& \(N_{\text{pseudo}}\) & ONMI (\%) 
& \(N_{\text{pseudo}}\) & ONMI (\%) \\
\midrule
Initial Training  & 6484 & 50.1 ± 3.8 & 4131 & 50.3 ± 5.5 & 2116 & 55.9 ± 4.7 & 14302 & 53.6 ± 2.0 \\
Refined Training & 22108 & \textbf{59.2 ± 1.4} & 13791 & \textbf{61.9 ± 1.5} & 9403 & \textbf{65.8 ± 1.1} & 39047 & \textbf{59.3 ± 1.5} \\
\bottomrule
\end{tabular}%
}
\caption{Comparison results between first and second rounds on MAG datasets.}
\label{tab_pseudo_generation_MAG}
\end{table}

\textbf{Exp-4: Effectiveness of Refined Training.} 
Tables~\ref{tab_pseudo_generation_FB} and~\ref{tab_pseudo_generation_MAG} present the comparison results between initial and refined training.
Specifically, we report the number of pseudo-labels \(N_{\text{pseudo}}\) used in each training round and the corresponding performance.
WOCD with refined training consistently achieves superior accuracy and stability across most datasets, with particularly notable improvements on the four large-scale MAG datasets.
The results demonstrate that updating pseudo-labels using a well-trained model after the first round allows the supervision to cover a broader range of nodes, enhancing performance and robustness.
This helps mitigate the noise introduced by the randomness of sampled nodes and the heterogeneity in the weak clique generation process.

\subsection{Parameter Analysis}
In this subsection, we investigate the impact of the sampling ratio \(\rho\) on all semi-supervised baseline models in comparison with WOCD.
Specifically, we vary \(\rho\) from 5\% to 20\% with a step size of 5\% to assess whether WOCD can maintain competitive detection performance under different levels of prior supervision.

\begin{figure}[H]
    \centering
    \includegraphics[width=\textwidth]{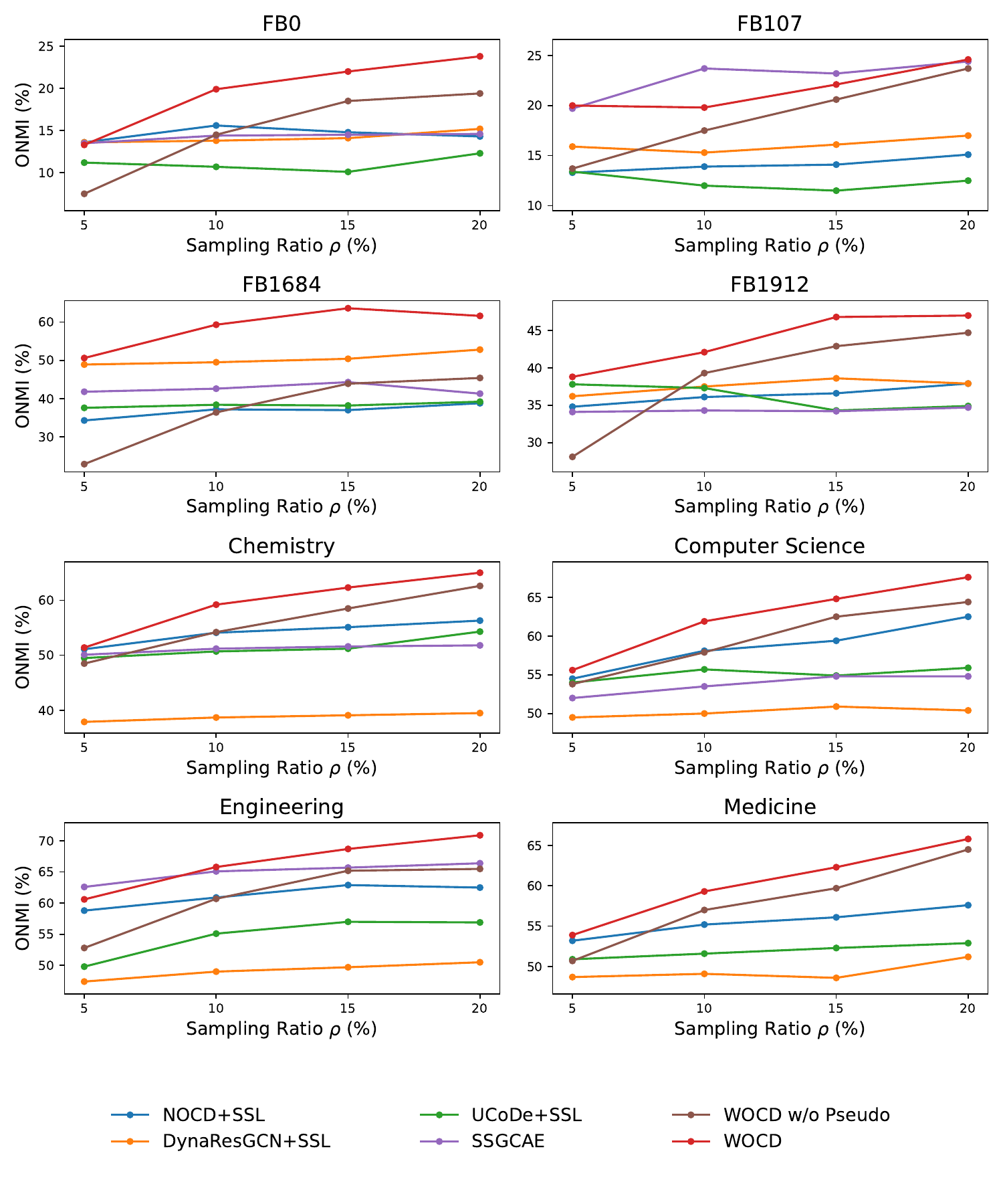}
    \caption{Impact of different levels of prior supervision on model performance.}
    \label{fig_rou_full}
\end{figure}

As shown in Figure~\ref{fig_rou_full}, WOCD consistently outperforms all other semi-supervised baselines across all supervision levels, especially when more prior information is available.
In addition, compared with its variant WOCD w/o Pseudo, WOCD demonstrates clear advantages under all settings. 
These results highlight the effectiveness of utilizing weak cliques to construct informative pseudo-labels, enabling WOCD to better adapt to varying degrees of prior supervision and consistently achieve superior performance.

\section{Conclusion}
\label{sec_conclusion}

In this paper, we propose a Weak-clique-based Overlapping Community Detection method, WOCD, which effectively integrates link, attribute, and prior information. 
WOCD leverages a novel semi-supervised framework with pseudo-labels to enhance generalization in real-world applications. 
To improve model stability, WOCD assigns high-quality pseudo-labels to certain nodes by identifying weak cliques before training. 
Furthermore, WOCD incorporates a lightweight Graph Transformer with GNN to expand its receptive field, achieving higher performance while maintaining efficiency. 
Compared to six existing OCD models, including both semi-supervised and unsupervised approaches, WOCD achieves the highest detection accuracy on 7 out of 8 datasets. 
Ablation studies further validate the effectiveness of the pseudo-labels, weak cliques, and the Graph Transformer introduced in WOCD.

\newpage
\appendix
\section{Architectural Details}
\label{app_architectural_details}


\subsection{Weak Clique Identification}
\begin{figure}[!htb]  
  \centering  
  \includegraphics[width=\textwidth]{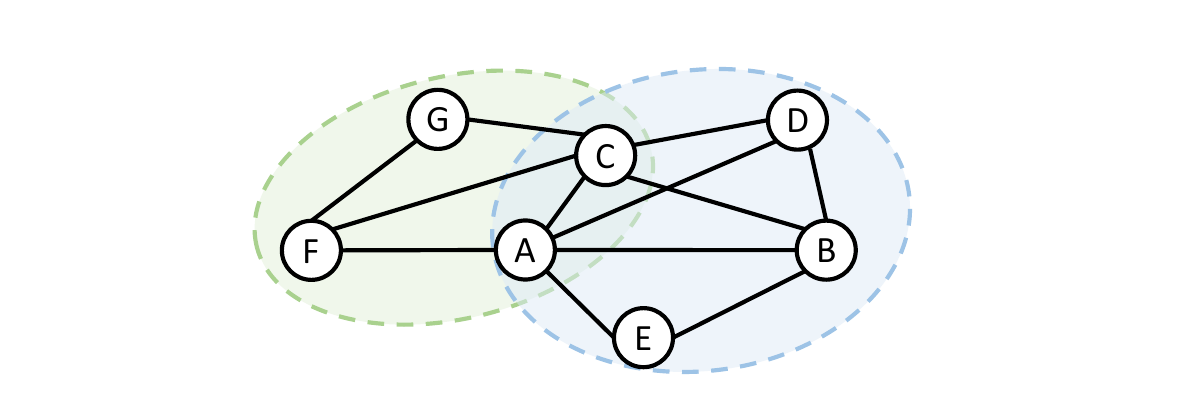}  
  \caption{An example of weak clique identification}
  \label{fig_weak_clique_example}  
\end{figure}

Figure \ref{fig_weak_clique_example} provides an example of identifying weak cliques. 
The graph contains two overlapping communities. 
Community 1 consists of nodes \(\{A, B,\allowbreak C, D, E\}\), while Community 2 consists of nodes \(\{A, C, F, G\}\). 
Based on the priority calculation, the nodes are ranked in descending order as follows: \(C = A > B > D > F > G = E\). 
Using the Salton index for weak clique identification, the initial pairs of nodes selected as starting nodes for weak cliques are: \((C, F)\), \((A, B)\), \((D, B)\), \((G, F)\), and \((E, B)\).
As a result, two weak cliques are identified: \(\{A, C, F, G\}\) and \(\{A, B, C, D, E\}\).

\section{Baseline Methods and Implementation}
\label{app_baselines}

\subsection{Baselines}
We evaluate six baseline methods, which can be divided into three categories:

Unsupervised non-parametric OCD Methods:

\begin{itemize}

\item \textbf{W-CPM~\cite{8047969}:} 
W-CPM is an overlapping community detection method based on the clique percolation method (CPM).
It reduces CPM's high computational complexity by identifying and merging weak cliques based on their overlap, making it particularly effective for detecting communities in large-scale networks.

\item \textbf{GREESE~\cite{10.1007/s00607-021-00948-4}:} 
GREESE is a seed-coupled expansion method for overlapping community detection. 
It can be divided into four primary steps: selecting community cores, expanding by adding core neighbors, assigning isolated nodes, and merging smaller communities.

\end{itemize}

Unsupervised parametric OCD methods:

\begin{itemize}
\item \textbf{NOCD~\cite{shchur2019overlappingcommunitydetectiongraph}:} 
NOCD is an unsupervised learning model. 
The main idea of NOCD is to use the Bernoulli-Poisson (BP) model to simulate the graph generation process, and employ a graph neural network (GNN) to predict the community affiliation of each node.

\item \textbf{DynaResGCN~\cite{muttakin2023overlappingcommunitydetectionusing}:} 
DynaResGCN is an unsupervised learning model. 
It combines dynamic residual Graph Convolutional Network (GCN) with the BP model based on NOCD, resulting in improved detection performance.

\item \textbf{UCoDe~\cite{moradan2023ucodeunifiedcommunitydetection}:} UCoDe is an unsupervised community detection model that supports both non-overlapping and overlapping communities. 
It leverages a GCN to optimize a novel contrastive loss function grounded in modularity.
\end{itemize}

Semi-supervised parametric OCD methods:
\begin{itemize}

\item \textbf{SSGCAE~\cite{HE20221464}:} 
SSGCAE is a semi-supervised learning model that combines link and attribute information using a graph convolutional autoencoder. 
It further improves community detection performance through modularity maximization and semi-supervised learning.

\end{itemize}

To address the lack of semi-supervised baselines, we extend three existing unsupervised methods (NOCD, DynaResGCN, and UCoDe) by incorporating semi-supervised losses, following the same protocol used in SSGCAE for a fair comparison.
To verify the effectiveness and stability of our weak clique-based pseudo-labeling strategy, we construct \textbf{WOCD w/o Pseudo}, a variant of WOCD with the same network architecture but trained solely on ground truth labels. 
Specifically, we disable the pseudo-label loss term by setting \(\lambda_2 = 0\) in Equation~\ref{equ_loss_function}.

\subsection{Implementation}
For all baseline methods, we use the hyperparameters provided by the original authors when available. Otherwise, we perform the grid search to identify the best hyperparameter settings.
Furthermore, to address the lack of semi-supervised baselines, we incorporate a semi-supervised loss into three unsupervised parametric methods.
Specifically, we follow the semi-supervised module in SSGCAE, which applies a cross-entropy loss $\mathcal{L}_{\text{SS}}$ between the sampled ground truth community affiliation matrix $\mathbf{C}_{\text{sampled}}$ and the model-predicted affiliation matrix $\mathbf{C}_{\text{pred}}$. The final objective is formulated as:
\begin{equation}
\label{equ_semi_loss_function}
\mathcal{L} = \mathcal{L}_{\text{base}} + \alpha \mathcal{L}_{\text{SS}}
\end{equation}
where \(\mathcal{L}_{\text{base}}\) denotes the original loss of the unsupervised model, and \(\alpha\) is a balancing hyperparameter. Following SSGCAE, we set \(\alpha\) as a fixed hyperparameter and perform the grid search to determine the optimal value for each baseline on every dataset, ensuring a fair comparison.

\bibliographystyle{elsarticle-num}
\bibliography{mybibliography}  

\begin{thebibliography}{10}
\expandafter\ifx\csname url\endcsname\relax
  \def\url#1{\texttt{#1}}\fi
\expandafter\ifx\csname urlprefix\endcsname\relax\def\urlprefix{URL }\fi
\expandafter\ifx\csname href\endcsname\relax
  \def\href#1#2{#2} \def\path#1{#1}\fi

\bibitem{Chen_2010}
P.~Chen, S.~Redner, \href{http://dx.doi.org/10.1016/j.joi.2010.01.001}{Community structure of the physical review citation network}, Journal of Informetrics 4~(3) (2010) 278–290.
\newblock \href {https://doi.org/10.1016/j.joi.2010.01.001} {\path{doi:10.1016/j.joi.2010.01.001}}.
\newline\urlprefix\url{http://dx.doi.org/10.1016/j.joi.2010.01.001}

\bibitem{NICOLINI201728}
C.~Nicolini, C.~Bordier, A.~Bifone, \href{https://www.sciencedirect.com/science/article/pii/S1053811916306449}{Community detection in weighted brain connectivity networks beyond the resolution limit}, NeuroImage 146 (2017) 28--39.
\newblock \href {https://doi.org/https://doi.org/10.1016/j.neuroimage.2016.11.026} {\path{doi:https://doi.org/10.1016/j.neuroimage.2016.11.026}}.
\newline\urlprefix\url{https://www.sciencedirect.com/science/article/pii/S1053811916306449}

\bibitem{10.1007/s11280-021-00966-4}
M.~Zarezadeh, E.~Nourani, A.~Bouyer, \href{https://doi.org/10.1007/s11280-021-00966-4}{Dpnlp: distance based peripheral nodes label propagation algorithm for community detection in social networks}, World Wide Web 25~(1) (2022) 73–98.
\newblock \href {https://doi.org/10.1007/s11280-021-00966-4} {\path{doi:10.1007/s11280-021-00966-4}}.
\newline\urlprefix\url{https://doi.org/10.1007/s11280-021-00966-4}

\bibitem{Fortunato_2016}
S.~Fortunato, D.~Hric, \href{http://dx.doi.org/10.1016/j.physrep.2016.09.002}{Community detection in networks: A user guide}, Physics Reports 659 (2016) 1–44.
\newblock \href {https://doi.org/10.1016/j.physrep.2016.09.002} {\path{doi:10.1016/j.physrep.2016.09.002}}.
\newline\urlprefix\url{http://dx.doi.org/10.1016/j.physrep.2016.09.002}

\bibitem{liu2024communitydetectiongraphneural}
C.~Liu, Y.~Han, H.~Xu, S.~Yang, K.~Wang, Y.~Su, \href{https://arxiv.org/abs/2401.02542}{A community detection and graph neural network based link prediction approach for scientific literature} (2024).
\newblock \href {http://arxiv.org/abs/2401.02542} {\path{arXiv:2401.02542}}.
\newline\urlprefix\url{https://arxiv.org/abs/2401.02542}

\bibitem{Wu_2022}
X.~Wu, Y.~Xiong, Y.~Zhang, Y.~Jiao, C.~Shan, Y.~Sun, Y.~Zhu, P.~S. Yu, \href{http://dx.doi.org/10.1145/3534678.3539370}{Clare: A semi-supervised community detection algorithm}, in: Proceedings of the 28th ACM SIGKDD Conference on Knowledge Discovery and Data Mining, KDD ’22, ACM, 2022, p. 2059–2069.
\newblock \href {https://doi.org/10.1145/3534678.3539370} {\path{doi:10.1145/3534678.3539370}}.
\newline\urlprefix\url{http://dx.doi.org/10.1145/3534678.3539370}

\bibitem{yang2012structureoverlapscommunitiesnetworks}
J.~Yang, J.~Leskovec, \href{https://arxiv.org/abs/1205.6228}{Structure and overlaps of communities in networks} (2012).
\newblock \href {http://arxiv.org/abs/1205.6228} {\path{arXiv:1205.6228}}.
\newline\urlprefix\url{https://arxiv.org/abs/1205.6228}

\bibitem{Palla_2005}
G.~Palla, I.~Derényi, I.~Farkas, T.~Vicsek, \href{http://dx.doi.org/10.1038/nature03607}{Uncovering the overlapping community structure of complex networks in nature and society}, Nature 435~(7043) (2005) 814–818.
\newblock \href {https://doi.org/10.1038/nature03607} {\path{doi:10.1038/nature03607}}.
\newline\urlprefix\url{http://dx.doi.org/10.1038/nature03607}

\bibitem{Gregory_2010}
S.~Gregory, \href{https://dx.doi.org/10.1088/1367-2630/12/10/103018}{Finding overlapping communities in networks by label propagation}, New Journal of Physics 12~(10) (2010) 103018.
\newblock \href {https://doi.org/10.1088/1367-2630/12/10/103018} {\path{doi:10.1088/1367-2630/12/10/103018}}.
\newline\urlprefix\url{https://dx.doi.org/10.1088/1367-2630/12/10/103018}

\bibitem{6341758}
V.~Y. Tan, C.~Févotte, Automatic relevance determination in nonnegative matrix factorization with the /spl beta/-divergence, IEEE Transactions on Pattern Analysis and Machine Intelligence 35~(7) (2013) 1592--1605.
\newblock \href {https://doi.org/10.1109/TPAMI.2012.240} {\path{doi:10.1109/TPAMI.2012.240}}.

\bibitem{PhysRevE.83.066114}
I.~Psorakis, S.~Roberts, M.~Ebden, B.~Sheldon, \href{https://link.aps.org/doi/10.1103/PhysRevE.83.066114}{Overlapping community detection using bayesian non-negative matrix factorization}, Phys. Rev. E 83 (2011) 066114.
\newblock \href {https://doi.org/10.1103/PhysRevE.83.066114} {\path{doi:10.1103/PhysRevE.83.066114}}.
\newline\urlprefix\url{https://link.aps.org/doi/10.1103/PhysRevE.83.066114}

\bibitem{8970691}
F.~Ye, C.~Chen, Z.~Zheng, R.-H. Li, J.~X. Yu, Discrete overlapping community detection with pseudo supervision, in: 2019 IEEE International Conference on Data Mining (ICDM), 2019, pp. 708--717.
\newblock \href {https://doi.org/10.1109/ICDM.2019.00081} {\path{doi:10.1109/ICDM.2019.00081}}.

\bibitem{shchur2019overlappingcommunitydetectiongraph}
O.~Shchur, S.~Günnemann, \href{https://arxiv.org/abs/1909.12201}{Overlapping community detection with graph neural networks} (2019).
\newblock \href {http://arxiv.org/abs/1909.12201} {\path{arXiv:1909.12201}}.
\newline\urlprefix\url{https://arxiv.org/abs/1909.12201}

\bibitem{muttakin2023overlappingcommunitydetectionusing}
M.~N. Muttakin, M.~I. Hossain, M.~S. Rahman, \href{https://arxiv.org/abs/2210.11174}{Overlapping community detection using dynamic dilated aggregation in deep residual gcn} (2023).
\newblock \href {http://arxiv.org/abs/2210.11174} {\path{arXiv:2210.11174}}.
\newline\urlprefix\url{https://arxiv.org/abs/2210.11174}

\bibitem{HE20221464}
C.~He, Y.~Zheng, J.~Cheng, Y.~Tang, G.~Chen, H.~Liu, \href{https://www.sciencedirect.com/science/article/pii/S0020025522007253}{Semi-supervised overlapping community detection in attributed graph with graph convolutional autoencoder}, Information Sciences 608 (2022) 1464--1479.
\newblock \href {https://doi.org/https://doi.org/10.1016/j.ins.2022.07.036} {\path{doi:https://doi.org/10.1016/j.ins.2022.07.036}}.
\newline\urlprefix\url{https://www.sciencedirect.com/science/article/pii/S0020025522007253}

\bibitem{LIU2017304}
X.~Liu, W.~Wang, D.~He, P.~Jiao, D.~Jin, C.~V. Cannistraci, \href{https://www.sciencedirect.com/science/article/pii/S0020025516318709}{Semi-supervised community detection based on non-negative matrix factorization with node popularity}, Information Sciences 381 (2017) 304--321.
\newblock \href {https://doi.org/https://doi.org/10.1016/j.ins.2016.11.028} {\path{doi:https://doi.org/10.1016/j.ins.2016.11.028}}.
\newline\urlprefix\url{https://www.sciencedirect.com/science/article/pii/S0020025516318709}

\bibitem{6985550}
L.~Yang, X.~Cao, D.~Jin, X.~Wang, D.~Meng, A unified semi-supervised community detection framework using latent space graph regularization, IEEE Transactions on Cybernetics 45~(11) (2015) 2585--2598.

\bibitem{alon2021bottleneckgraphneuralnetworks}
U.~Alon, E.~Yahav, \href{https://arxiv.org/abs/2006.05205}{On the bottleneck of graph neural networks and its practical implications} (2021).
\newblock \href {http://arxiv.org/abs/2006.05205} {\path{arXiv:2006.05205}}.
\newline\urlprefix\url{https://arxiv.org/abs/2006.05205}

\bibitem{wu2022representing}
Z.~Wu, P.~Jain, M.~A. Wright, A.~Mirhoseini, J.~E. Gonzalez, I.~Stoica, \href{https://arxiv.org/abs/2201.08821}{Representing long-range context for graph neural networks with global attention} (2022).
\newblock \href {http://arxiv.org/abs/2201.08821} {\path{arXiv:2201.08821}}.
\newline\urlprefix\url{https://arxiv.org/abs/2201.08821}

\bibitem{zhao2020pairnorm}
L.~Zhao, L.~Akoglu, \href{https://arxiv.org/abs/1909.12223}{Pairnorm: Tackling oversmoothing in gnns} (2020).
\newblock \href {http://arxiv.org/abs/1909.12223} {\path{arXiv:1909.12223}}.
\newline\urlprefix\url{https://arxiv.org/abs/1909.12223}

\bibitem{chen2019measuring}
D.~Chen, Y.~Lin, W.~Li, P.~Li, J.~Zhou, X.~Sun, \href{https://arxiv.org/abs/1909.03211}{Measuring and relieving the over-smoothing problem for graph neural networks from the topological view} (2019).
\newblock \href {http://arxiv.org/abs/1909.03211} {\path{arXiv:1909.03211}}.
\newline\urlprefix\url{https://arxiv.org/abs/1909.03211}

\bibitem{DANESHFAR2024108215}
F.~Daneshfar, S.~Soleymanbaigi, P.~Yamini, M.~S. Amini, \href{https://www.sciencedirect.com/science/article/pii/S0952197624003737}{A survey on semi-supervised graph clustering}, Engineering Applications of Artificial Intelligence 133 (2024) 108215.
\newblock \href {https://doi.org/https://doi.org/10.1016/j.engappai.2024.108215} {\path{doi:https://doi.org/10.1016/j.engappai.2024.108215}}.
\newline\urlprefix\url{https://www.sciencedirect.com/science/article/pii/S0952197624003737}

\bibitem{SHI2013394}
C.~Shi, Y.~Cai, D.~Fu, Y.~Dong, B.~Wu, \href{https://www.sciencedirect.com/science/article/pii/S0169023X13000499}{A link clustering based overlapping community detection algorithm}, Data \& Knowledge Engineering 87 (2013) 394--404.
\newblock \href {https://doi.org/https://doi.org/10.1016/j.datak.2013.05.004} {\path{doi:https://doi.org/10.1016/j.datak.2013.05.004}}.
\newline\urlprefix\url{https://www.sciencedirect.com/science/article/pii/S0169023X13000499}

\bibitem{Lancichinetti_2009}
A.~Lancichinetti, S.~Fortunato, J.~Kertész, \href{http://dx.doi.org/10.1088/1367-2630/11/3/033015}{Detecting the overlapping and hierarchical community structure in complex networks}, New Journal of Physics 11~(3) (2009) 033015.
\newblock \href {https://doi.org/10.1088/1367-2630/11/3/033015} {\path{doi:10.1088/1367-2630/11/3/033015}}.
\newline\urlprefix\url{http://dx.doi.org/10.1088/1367-2630/11/3/033015}

\bibitem{lee2010detectinghighlyoverlappingcommunity}
C.~Lee, F.~Reid, A.~McDaid, N.~Hurley, \href{https://arxiv.org/abs/1002.1827}{Detecting highly overlapping community structure by greedy clique expansion} (2010).
\newblock \href {http://arxiv.org/abs/1002.1827} {\path{arXiv:1002.1827}}.
\newline\urlprefix\url{https://arxiv.org/abs/1002.1827}

\bibitem{Ahn_2010}
Y.-Y. Ahn, J.~P. Bagrow, S.~Lehmann, \href{http://dx.doi.org/10.1038/nature09182}{Link communities reveal multiscale complexity in networks}, Nature 466~(7307) (2010) 761–764.
\newblock \href {https://doi.org/10.1038/nature09182} {\path{doi:10.1038/nature09182}}.
\newline\urlprefix\url{http://dx.doi.org/10.1038/nature09182}

\bibitem{Kumpula_2008}
J.~M. Kumpula, M.~Kivelä, K.~Kaski, J.~Saramäki, \href{http://dx.doi.org/10.1103/PhysRevE.78.026109}{Sequential algorithm for fast clique percolation}, Physical Review E 78~(2) (Aug. 2008).
\newblock \href {https://doi.org/10.1103/physreve.78.026109} {\path{doi:10.1103/physreve.78.026109}}.
\newline\urlprefix\url{http://dx.doi.org/10.1103/PhysRevE.78.026109}

\bibitem{8944086}
J.~Ma, J.~Fan, Local optimization for clique-based overlapping community detection in complex networks, IEEE Access 8 (2020) 5091--5103.
\newblock \href {https://doi.org/10.1109/ACCESS.2019.2962751} {\path{doi:10.1109/ACCESS.2019.2962751}}.

\bibitem{6968420}
S.~Maity, S.~K. Rath, Extended clique percolation method to detect overlapping community structure, in: 2014 International Conference on Advances in Computing, Communications and Informatics (ICACCI), 2014, pp. 31--37.
\newblock \href {https://doi.org/10.1109/ICACCI.2014.6968420} {\path{doi:10.1109/ICACCI.2014.6968420}}.

\bibitem{8047969}
X.~Zhang, C.~Wang, Y.~Su, L.~Pan, H.-F. Zhang, A fast overlapping community detection algorithm based on weak cliques for large-scale networks, IEEE Transactions on Computational Social Systems 4~(4) (2017) 218--230.
\newblock \href {https://doi.org/10.1109/TCSS.2017.2749282} {\path{doi:10.1109/TCSS.2017.2749282}}.

\bibitem{moradan2023ucodeunifiedcommunitydetection}
A.~Moradan, A.~Draganov, D.~Mottin, I.~Assent, \href{https://arxiv.org/abs/2112.14822}{Ucode: Unified community detection with graph convolutional networks} (2023).
\newblock \href {http://arxiv.org/abs/2112.14822} {\path{arXiv:2112.14822}}.
\newline\urlprefix\url{https://arxiv.org/abs/2112.14822}

\bibitem{10.5555/576628}
G.~Salton, M.~J. McGill, Introduction to Modern Information Retrieval, McGraw-Hill, Inc., USA, 1986.

\bibitem{kipf2017graphconvolutional}
T.~N. Kipf, M.~Welling, \href{https://arxiv.org/abs/1609.02907}{Semi-supervised classification with graph convolutional networks} (2017).
\newblock \href {http://arxiv.org/abs/1609.02907} {\path{arXiv:1609.02907}}.
\newline\urlprefix\url{https://arxiv.org/abs/1609.02907}

\bibitem{wu2024sgformersimplifyingempoweringtransformers}
Q.~Wu, W.~Zhao, C.~Yang, H.~Zhang, F.~Nie, H.~Jiang, Y.~Bian, J.~Yan, \href{https://arxiv.org/abs/2306.10759}{Sgformer: Simplifying and empowering transformers for large-graph representations} (2024).
\newblock \href {http://arxiv.org/abs/2306.10759} {\path{arXiv:2306.10759}}.
\newline\urlprefix\url{https://arxiv.org/abs/2306.10759}

\bibitem{10.1007/s00607-021-00948-4}
K.~Asmi, D.~Lotfi, A.~Abarda, \href{https://doi.org/10.1007/s00607-021-00948-4}{The greedy coupled-seeds expansion method for the overlapping community detection in social networks}, Computing 104~(2) (2022) 295–313.
\newblock \href {https://doi.org/10.1007/s00607-021-00948-4} {\path{doi:10.1007/s00607-021-00948-4}}.
\newline\urlprefix\url{https://doi.org/10.1007/s00607-021-00948-4}

\bibitem{10.1145/2556612}
J.~Mcauley, J.~Leskovec, \href{https://doi.org/10.1145/2556612}{Discovering social circles in ego networks}, ACM Trans. Knowl. Discov. Data 8~(1) (feb 2014).
\newblock \href {https://doi.org/10.1145/2556612} {\path{doi:10.1145/2556612}}.
\newline\urlprefix\url{https://doi.org/10.1145/2556612}

\bibitem{9722614}
S.~Yuan, C.~Wang, Q.~Jiang, J.~Ma, Community detection with graph neural network using markov stability, in: 2022 International Conference on Artificial Intelligence in Information and Communication (ICAIIC), 2022, pp. 437--442.
\newblock \href {https://doi.org/10.1109/ICAIIC54071.2022.9722614} {\path{doi:10.1109/ICAIIC54071.2022.9722614}}.

\bibitem{wang2024efficientunsupervisedcommunitysearch}
J.~Wang, K.~Wang, X.~Lin, W.~Zhang, Y.~Zhang, \href{https://arxiv.org/abs/2403.18869}{Efficient unsupervised community search with pre-trained graph transformer} (2024).
\newblock \href {http://arxiv.org/abs/2403.18869} {\path{arXiv:2403.18869}}.
\newline\urlprefix\url{https://arxiv.org/abs/2403.18869}

\bibitem{wang2024neuralattributedcommunitysearch}
J.~Wang, K.~Wang, X.~Lin, W.~Zhang, Y.~Zhang, \href{https://arxiv.org/abs/2403.18874}{Neural attributed community search at billion scale} (2024).
\newblock \href {http://arxiv.org/abs/2403.18874} {\path{arXiv:2403.18874}}.
\newline\urlprefix\url{https://arxiv.org/abs/2403.18874}

\bibitem{kingma2017adammethodstochasticoptimization}
D.~P. Kingma, J.~Ba, \href{https://arxiv.org/abs/1412.6980}{Adam: A method for stochastic optimization} (2017).
\newblock \href {http://arxiv.org/abs/1412.6980} {\path{arXiv:1412.6980}}.
\newline\urlprefix\url{https://arxiv.org/abs/1412.6980}

\bibitem{mcdaid2013normalizedmutualinformationevaluate}
A.~F. McDaid, D.~Greene, N.~Hurley, \href{https://arxiv.org/abs/1110.2515}{Normalized mutual information to evaluate overlapping community finding algorithms} (2013).
\newblock \href {http://arxiv.org/abs/1110.2515} {\path{arXiv:1110.2515}}.
\newline\urlprefix\url{https://arxiv.org/abs/1110.2515}

\end{thebibliography}

\end{document}